\documentstyle[11pt,newpasp,twoside,epsf]{article}
\def\edcomment#1{\iffalse\marginpar{\raggedright\sl#1\/}\else\relax\fi}
\reversemarginpar

\begin{document}
\title{The Impact of Galaxies on their Environment from Observations
of Gravitationally Lensed QSOs}
\author{Michael Rauch}
\affil{Carnegie Observatories, 813 Santa Barbara St., Pasadena, CA 91101}


\begin{abstract}
Observations of absorption systems in close, multiple lines of sight to
gravitationally lensed QSOs can be used to infer the density
fluctuations and motions of the gas clouds giving rise to the Lyman
$\alpha$ forest phenomenon and to QSO metal absorption systems.  We
describe a survey of lensed QSOs with the Keck HIRES instrument and
argue that one can derive limits on the frequency and impact of
hydrodynamical disturbances inflicted by galaxies on the surrounding
gas from such data. We discuss differences between the kinematic
properties of low density unsaturated Ly$\alpha$ forest absorption
systems, high ionization CIV absorption systems, and low ionization gas
visible in (e.g.) SiII and CII.  The general intergalactic medium (as
seen in the Ly$\alpha$ forest) shows very little turbulence, but the
presumably denser CIV systems exhibit evidence of having been stirred
repeatedly (by winds ?) in the past on time scales similar to those
governing stellar feedback and possibly galaxy mergers.  The quiescence
of the low density IGM can be used to put upper limits on the incidence
and energetics of galactic winds on a cosmological scale.

\end{abstract}

\section{Introduction}

Historically, studying common absorption systems in the spectra of
lensed QSOs was the first astrophysical application of gravitational
lensing; in fact it was the presence of common absorption in the two
images of 0957+561 which provided strong evidence for the lens nature
of the first such system discovered (Walsh, Carswell \& Weymann, 1979).
Ray Weymann, in a pattern that some of the participants in this
workshop may recognize, has made seminal contributions to this subject,
including being the co-discoverer of the first three lensed QSOs (Walsh
et al 1979; Weymann et al 1979; Weymann et al 1980; Weedman et al
1982), then to largely leave the topic for others to sort out the
details - and we still are trying to after more than 20 years. There
certainly has been substantial progress - I only mention here the
important work by Peter Young (e.g., Young et al 1981a; 1981b) and by
Alain Smette (e.g., Smette et al 1992; 1995) and their collaborators.
But good spectroscopic data have been hard to obtain. The lensed images
of QSOs mostly are too faint in the optical to be examined with a high
resolution spectrograph; and to measure typical velocity differences in
gas clouds between two lensed lines of sight (i.e., proper separations
of mostly less than a few kpcs) a resolution of a few kms$^{-1}$ is
required.  The advent of 8-10m class telescopes finally heralded in the
post-heroic age, where such high resolution QSO spectra could be
obtained in a matter of hours, not days.  In the present contribution I
describe a survey my collaborators Wal Sargent, Tom Barlow, and I began
with the HIRES instrument (Vogt et al 1984) on Keck I a few years ago.
HIRES was the first instrument to allow high resolution (FWHM $\sim 6$
kms$^{-1}$) spectroscopy down to about 19th magnitude in the optical
wavelength range, so several of the brightest known lenses (among them
UM673 (Surdej et al. 1988; Smette et al 1992); HE1104-1805 (Wisotzki et
al 1993; Smette et al 1995) and Q1422+213 (Patnaik et al. 1992;
Bechtold \& Yee 1995; Petry et al 1998)) and a number of other objects
came within reach of having at least two images each bright enough to
be examined at high resolution.  The principle of using gravitational
lenses to study the properties of absorption systems is illustrated in
figure 1.

\begin{figure}
\plotfiddle{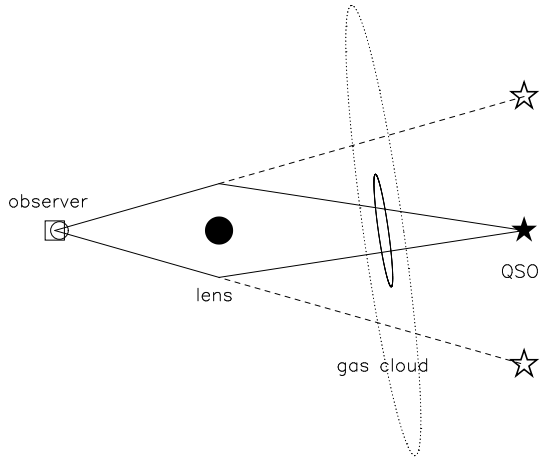}{3.cm}{-90.}{35}{35}{-140}{120}
\vskip2.5cm
\caption{\small 
a galaxy lens (black dot) produces two displaced images (open star
symbols) of a background QSO, which, without the lensing effect would
appear as a single point source (filled star). Imagine that
a gas cloud (ellipse with solid outline) is situated between the
lens and the QSO causing absorption in one or both lines of sight. 
The angular magnification produced by the lens makes
this cloud appear as if it were much bigger in projection on the sky
(dotted ellipse). The higher the redshift of the cloud the smaller is
the separation between the lines of sight intersecting it. Thus the
combination of ground-based telescope and gravitational lens works like
a gigantic microscope for the matter intervening between the lens and
the QSO redshift. In practice it has been possible to study QSO
absorption systems at transverse line of sight separations of a few
tens of parsecs at redshifts 2-4.  }
\end{figure}

The differences between the absorption pattern in two adjacent lines of
sight can be characterized in a variety of ways. If one thinks of the
gas clouds in terms of coherent objects it makes sense to measure the
optical depth or column density differences between the lines of sight,
to determine the scale over which the gas densities vary and to get an
idea of the cloud sizes (if the transverse beam separation is wide
enough to sample these).  Differences across the lines of sight of the
velocities projected along the lines of sight (i.e. the velocity shear)
provide clues to turbulence and systematic motion (e.g., rotation,
expansion) in the gas.  We have applied the relevant techniques to
various classes of absorption systems in the abovementioned QSO
spectra and describe the results below for several astrophysical environments
in the order of increasing gas density.

\begin{figure}[t]
\plotfiddle{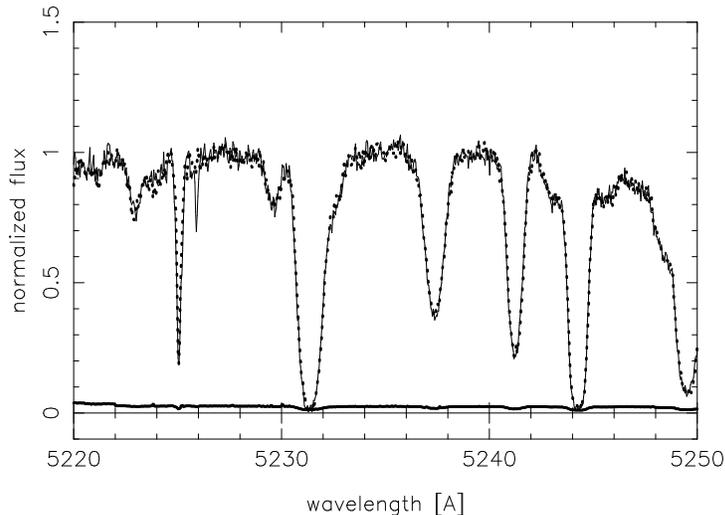}{4cm}{-90.}{40}{40}{-150}{170}
\vskip1.7cm
\caption{\small Enlargement of part of the spectrum of Q1422+231. The spectra of the A
(solid line) and C (dotted line) images are plotted on top of each
other.  There is hardly any difference between the spectra, with the
exception of a narrow line (which is largely absent in the spectrum of
the C image) and some lower column density fluctuations near 5226 \AA .
This line and the sharp stronger one at 5225 \AA\ can be identified
with a SiIV 1394\AA\ interloper from an absorption system at
$z=2.74889$. The other absorption features are plausibly attributed to
HI Ly$\alpha$.  The slowly varying function plotted at the bottom is the
$1-\sigma$ error of image A.  }
\end{figure}

\section{The Lyman $\alpha$ forest}

For practical purposes we restrict the term `Lyman $\alpha$ forest'
here to refer to absorption systems with unsaturated Ly$\alpha$ lines
only.  This distinction is based on the prejudice that such lines are
mostly associated with the general intergalactic medium, as opposed to
the interstellar medium or the halos of galaxies. Theoretical work and
cosmological hydro-simulations show that gas condensations giving rise
to such lines can be formed abundantly by gravitational collapse in
hierarchical scenarios of galaxy formation.  In the absence of other
dynamic agents capable of stirring them up (e.g., galactic winds and explosions)
these IGM clouds should be featureless on scales much smaller than the
Jeans length, because of the smoothing effects of thermal gas pressure.  This
prediction can be tested by measuring column density and velocity
differences between the lines of sight.  Our spectra of Q1422+231 A and
C ($z_{em}$ =3.62) intersect the IGM at mean redshift $<z>\sim 3.3$ and at a mean separation 110
$h_{50}^{-1}$ pc (for Ly$\alpha$ absorption lines between Ly$\beta$ and Ly$\alpha$
emission). Figure 2 shows a small section of the two spectra plotted on top 
of each other to demonstrate how small the differences between the two
spectra are over these scales. In contrast there is a sharp narrow SiIV interloper 
which is very different in the two lines of sight.  Metal absorption
lines often look different over these separations (they are usually associated
with strong, saturated Ly$\alpha$ lines) but the lower column density, mostly
unsaturated Ly$\alpha$ forest lines hardly show any difference,  reflecting qualitatively
different astrophysical environments in the two classes of absorption systems.

\subsection{By how much do the absorbers differ ?}

We have fitted the whole Ly$\alpha$ forest with Voigt profiles using the
fitting routine VPFIT (Carswell et al 1992) and determined the differences in column density
and velocity between the modelled absorption components of the two lines of sight.
For a subset of unsaturated Ly$\alpha$ lines with $12 < \log N < 14.13$, the observations
show that the RMS velocity differences  between the two lines of sight $A$ and $C$, over a mean separation of 0.110
h$_{50}^{-1}$ kpc, are less than
\begin{eqnarray}
\sqrt{\left<(v_A - v_C)^2\right>} \leq 0.4 {\rm kms}^{-1}.
\end{eqnarray}
The column density differences can be derived similarly. If the absorbing gas
is primarily formed by gravitational collapse, the HI column density differences
$\Delta N({\rm HI})$ can be translated into gas density differences $\Delta\rho$ using the tight correlation
$N({\rm HI})\propto{\rho^{\alpha}}$,
with $\alpha$ = 1.37 -1.5  (Schaye 2001), and the
variance of the logarithmic baryon density gradient becomes
\begin{eqnarray}
\left<\left(\Delta\log \rho\right)^2\right> \leq \alpha^{-2} \left<\left(\log N_A - \log N_C\right)^2\right>
\end{eqnarray}
or
\begin{eqnarray}
\sqrt{\left<\left(\Delta\log \rho\right)^2\right>} \leq 3\times10^{-2}
\end{eqnarray}
for the typical logarithmic change in density over 0.110
h$_{50}^{-1}$ kpc, i.e., the RMS fluctuations in the baryon density
are less than about 3 percent.

\subsection{What fraction of the Ly$\alpha$ forest has been disturbed ?}

The above analysis depends on the identification of absorption `lines'
and on measuring their properties. A more general approach may involve
just measuring the optical depths and comparing them for each pixel in
the spectra of adjacent lines of sight.  This way one could measure
(e.g.) the global fraction of the Ly$\alpha$ forest differing by more
than a certain amount in optical depth. The differences may be caused
by galaxies intersecting the line of sight, or by the presence of
galaxy winds or explosions stirring the IGM.  Defining the `disturbed
fraction of the Ly$\alpha$ forest', or the line of sight filling factor
$f_{LoS}$ as the fraction of the spectrum where the optical depths differ by more
than $\Delta \tau = |\tau_A - \tau_C| \leq 5\%$ between the lines of
sight, we measure $f_{LoS} \leq {0.23}$.  In other words: 77\% of the
Ly$\alpha$ forest spectrum do not  differ in optical depth by more than
than 5\% .

\subsection{Limits on the volume filling factor of winds}

Making some simplifying assumptions the above upper limit on the
line-of-sight filling factor $f_{LoS}$ can be translated into an
upper limit on the volume filling factor of the universe for the
disturbing processes, $f_v$. The relation between the two is of course
model-dependent. If we assume that the IGM is primarily disturbed by
strong spherically symmetric galactic winds and model these as
superbubbles escaping from galaxies into intergalactic space, we can
derive upper limits on $f_v$.  Applying the galactic super bubble model
of Mac Low \& McCray (1988) we can write the volume filling factor
$f_v$ for such winds as a function of the luminosity $L_{38}$ (in units
of $10^{38}$ ergs/s), the density of the ambient IGM $n_{-5}$ (in units of
$10^{-5}$ cm$^{-3}$), and the time elapsed between the start of the wind
and the observation. Some assumption must be made about the absorption
"footprint" of a wind-driven bubble. (Rauch,
Sargent \& Barlow 2001b).

\begin{figure}[t]
\plotfiddle{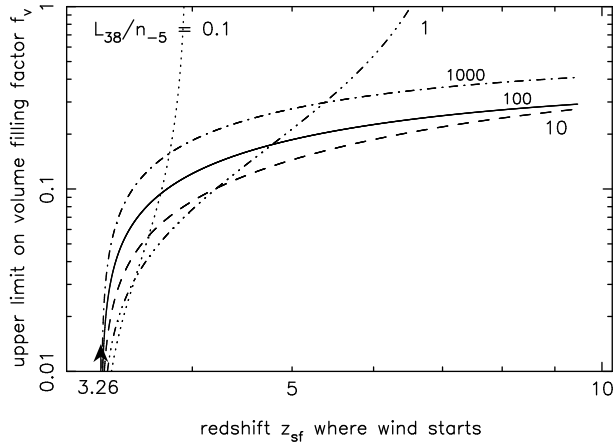}{3cm}{-90.}{35}{35}{-140}{130}
\vskip2.5cm
\caption{ \small   Upper limit on the volume filling factor for wind
blown bubbles. Strong winds with $L_{38}/n_{-5} \sim 100-1000$ can fill
a significant fraction of the volume of the universe (up to 40\% in the
extreme case, i.e, if they start as early as redshift 10, up to $\sim
18\% $ if they start at z=4), whereas winds with $L_{38}/n_{-5} \leq 1$
are essentially unconstrained if they occur before redshift 7.  }
\end{figure} 

The results are shown in fig.3. Strong winds $L_{38}/n_{-5} \sim
100-1000$ are well constrained because of their long survival times; a
volume filling factor of up to 40\% is consistent with our upper limit
on $f_{LoS}$, for the extreme case of $L_{38}/n_{-5} \sim 1000$ and a
wind starting as early as $z = 10$.  In contrast to that the absorption
signature of weak winds $L_{38}/n_{-5} \sim 0.1-1$ is wiped out much
sooner by gas pressure waves. Nevertheless,
weak winds cannot venture far beyond the filaments where their parent
galaxies reside and thus are unlikely to fill a substantial fraction of
the voids which make up most of the universe.

\section{Higher Gas Densities: Metal Absorption Systems}

Going to higher densities the Ly$\alpha$ forest lines become saturated
and invariably show simultaneous metal absorption lines, predominantly
the CIV doublet.  From the HI column densities it is to be expected
that this gas is residing much closer to galaxies than the typical
Ly$\alpha$ forest absorber. Nevertheless, photoionization calculations and
hydro-simulations of galaxy formation show that the underlying gas
density of CIV systems is mostly less than the virial density of a
galaxy, so a typical high redshift CIV system is not directly part of
a galaxy either. 

We have studied CIV systems in the three lensed QSOs
in regions longward of the Ly$\alpha$ emission line.  Because of their
lower source redshifts and higher lens redshifts, UM673 and HE1104-1805
provided lines of sight with wider separations than those of
Q1422+231.  Thus it was possible to probe CIV systems with separations
up to several kpc, and to investigate how the differences
between the column densities and column density weighted velocities
change with the separation between the lines of sight (Fig. 4).

\begin{figure}[t]
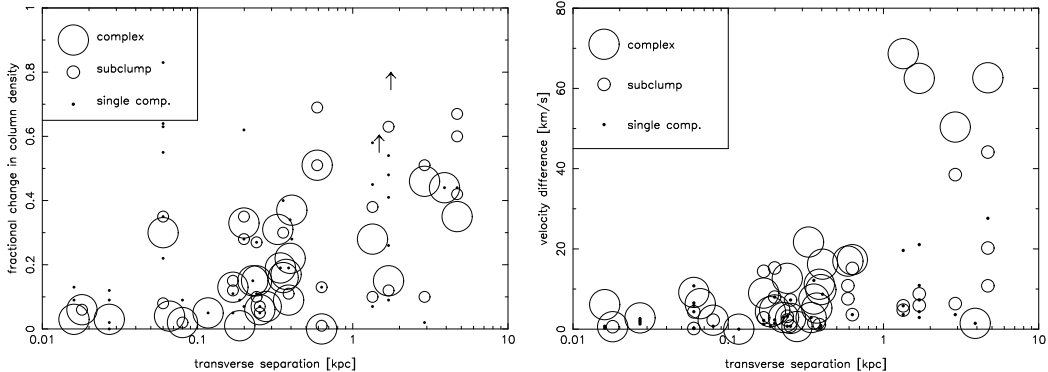

\plotfiddle{mrauch_4a.ps}{5cm}{-90.}{28}{28}{-10}{170}
\plotfiddle{mrauch_4b.ps}{5cm}{-90.}{28}{28}{-210}{325}
\vskip-5.5cm
\caption{ 
\small  Fractional difference between the column densities, $(N_A -
N_B)/({\rm max}(N_A,N_B)$ (left panel) and differences between the
column density weighted velocities $v_A - v_B$ (right panel) along the
line of sight, versus separation between the lines of sight, for a
sample of CIV systems from the three QSOs.  The different symbols
denote individual, "single" CIV components, subgroups of components,
and absorption systems as a whole ("complex").  The arrows show lower
limits (i.e., absorption at a given redshift only present in one of the
two lines of sight).}
\end{figure}

\subsection{Cloud sizes and turbulent energy}

A trend for the column densities and velocities to become more
discrepant with increasing beam separation $r$ is apparent in Fig.4.
The scale where this happens defines a minimum "size" for
the clouds.  Both, differences in column density at the 50\% level and
a substantial increase in the velocity scatter appear to occur at
separations on the order of a few hundred parsecs (albeit subject to a 
large scatter).

The existence of a minimum size for the CIV absorbers can be used to
get an estimate of the frequency with which the CIV gas has been stirred up
in the past. This is based on the idea that density and pressure
differences in a gas cloud left alone for a certain amount of time
without further disturbance from the outside would be smoothed by the
effects of thermal gas pressure, and the smoothing would occur on
spatial scales increasing with time.  Density gradients in the CIV gas
would be damped out by pressure waves propagating with the speed of
sound over a spatial distance  $r$ given by the product  of the  sound
crossing time $\tau_s$ and the sound speed, $c_s$.  If there is little
structure over a distance $r$, then there cannot have been a
hydrodynamic disturbance during the past \begin{eqnarray} \tau_s\sim
\frac{r}{c_s} \approx 1.4\times 10^7 \left(\frac{r}{300\mathrm
pc}\right) \left(\frac{c_s}{20 {\mathrm kms}^{-1}}\right)^{-1} {\mathrm
years}, \end{eqnarray} where a fiducial cloud size of 300 pc was
assumed.

Similarly, the velocity differences between the lines of sight can be
used to get an idea of the turbulent energy contents and rate of
energy input into the gas. Measuring the variance in the velocity difference
between the lines of sight A and B on spatial scale r, $v_r = \overline{(v_A(r)-v_B(r))^2}$ (details in
Rauch et al 2001a), and applying some simple consequences of the
Kolmogorov theory of turbulence, a crude estimate of the turbulent
energy input $\epsilon$ into the clouds is given by $\epsilon\sim
v_r^3/r \sim 10^{-3}{\mathrm cm}^2 s^{-3}$. This value is much smaller
than the energy input rate in galactic starforming regions
(e.g., Kaplan and Pikelner 1970), which is
another indication that, by observing high redshift CIV systems, we are
not usually looking straight through galaxies or starforming regions.
Nevertheless the turbulence observed is finite. We can again make the
argument that it cannot survive forever without further energy input.
If Kolmogorov-type conditions (esp. steady state energy input) were
applicable, the rate of energy input on large scales would equal the
rate of dissipation on the smallest scales.
Now, if the energy input were suddenly interrupted, the mean turbulent kinetic energy 
1/2 $<v^2>$ would be
turned into heat 
after a dissipation time 
\begin{eqnarray}
\tau_{\mathrm diss} \sim \frac{1}{2}\frac{<v^2>}{\epsilon} \sim 
9\times10^7{\mathrm years}.
\end{eqnarray}

This time scale is not dramatically different from the one obtained from the
minimum size of the clouds. Both estimates imply that CIV clouds are being
"stirred" on a time scale significantly smaller than a Hubble time.
Thus, even if the CIV gas represents, e.g., the relics of very early winds
deposited by a Population III phase of stellar nucleosynthesis they must
have been disturbed more recently, possibly by galactic winds or merger
events. There is evidence of recurrent star formation episodes in our
and other galaxies (e.g. Tomita, Tomita \& Saito 1996; Hirashita \&
Kamaya 2000;  Glazebrook et al 1999;
Rocha-Pinto et al 2000) and the dynamical state of CIV clouds may
reflect the effects of stellar feedback from these events upon the ambient
intergalactic medium.

\section{The ISM, Supernova Remnants, and Superbubbles} 

Occasionally, the QSO lines of sight must be looking through even
denser gas in regions directly related to a galactic environment, e.g.,
the interstellar medium in galactic disks, gaseous halos, or expanding
superbubbles. The problem is how to interpret the observed absorption
pattern in terms of the physical conditions in the underlying
astrophysical environment. Attempts at explaining low ionization
absorption systems globally by any particular astronomical scenario
have remained unconvincing (e.g., damped Lyman $\alpha$ systems as
galactic disks). The complexity of the ISM (as seen in our Galaxy)
clearly does not help here,
and it is possible that local gasdynamical processes (stellar
winds, supernova explosions) may play a more important dynamical role
than for example gravitational instability and thermal gas pressure,
which are thought to be the dominant agents in the intergalactic
medium.

Here we present some anectodal evidence that at least some of the low
ionization absorption systems have structure on very small (pc to kpc)
scales.  The first example is an object observed in the spectrum of QSO
1422+231 at a beam separation of only about 26 $h_{50}^{-1}$ pc (Rauch,
Sargent \& Barlow 1999).

\begin{figure}[t]
\plotfiddle{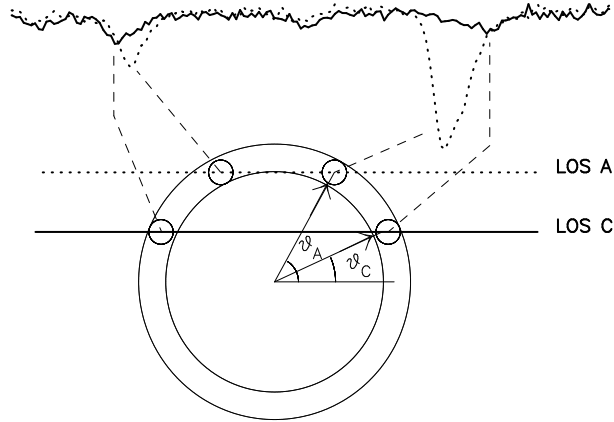}{4cm}{-90.}{35}{35}{-140}{140}
\vskip1.5cm
\caption{\small  z=3.538 CII absorption system in the $A$ (dotted spectrum)
and $C$ (solid spectrum) images of Q1422+231. The observed differences between the
spectra include larger column densities in the $A$ spectrum and a wider velocity
spacing between the two main absorption components in the $C$ spectrum.
Such a pattern could arise in an expanding shell of gas intersected by the two lines
of sight at different impact parameters. }
\end{figure} 

Although it is impossible to give a unique interpretation for any given
absorption system, the one discussed here (fig.5) has all the signs of arising
in an evolved interstellar medium of a galaxy at redshift 3.5.  
The object has strong low ionization lines from CII, SiII and OI. The ionization
state indicates a high density on the order of 0.1 - 2 cm$^{-3}$, just as in
the ISM of our galaxy. The metallicity is consistent with being solar. 
The pattern of column density and velocity differences can be explained
by the lines of sight intersecting an expanding shell of gas, possibly an
old supernova remnant or bubble, with a radius
$13 \leq R \leq 48$pc and an expansion velocity of $v_{\rm exp} \geq 98$kms$^{-1}$.
In any case, since the differences between the absorption systems arise over
only a few tens of parsecs the kinematics cannot reflect any
large scale bulk motion of the galaxy like rotation, but must be due to a local
process in the ISM.

The second case (fig.6) concerns an interesting low ionization system seen at
z=0.5656 in three lines of sight to the quadruply lensed QSO 2237+0305
(Huchra et al 1985; the fourth line of sight was not observed).  The
system, probably damped, shows strong MgII, FeII, and MgI lines, in
particular a central pair of lines, which seems to expand in
velocity space as one goes from $A$ to $C$ to $B$ in order of
increasing distance from $A$. This can be explained in a model where
all three lines of sight are looking through the same, expanding bubble of
gas, and each of the intersections with the bubble wall gives rise to an absorption
component. Again, such an explanation is not unique, of course. However, the
plausibility of an identification of MgII systems with expanding
superbubbles has also recently been advocated on other grounds by  Bond et al (2001).

\begin{figure}[t]
\plotfiddle{mrauch_6a.ps}{4cm}{-90.}{28}{28}{-220}{120}
\plotfiddle{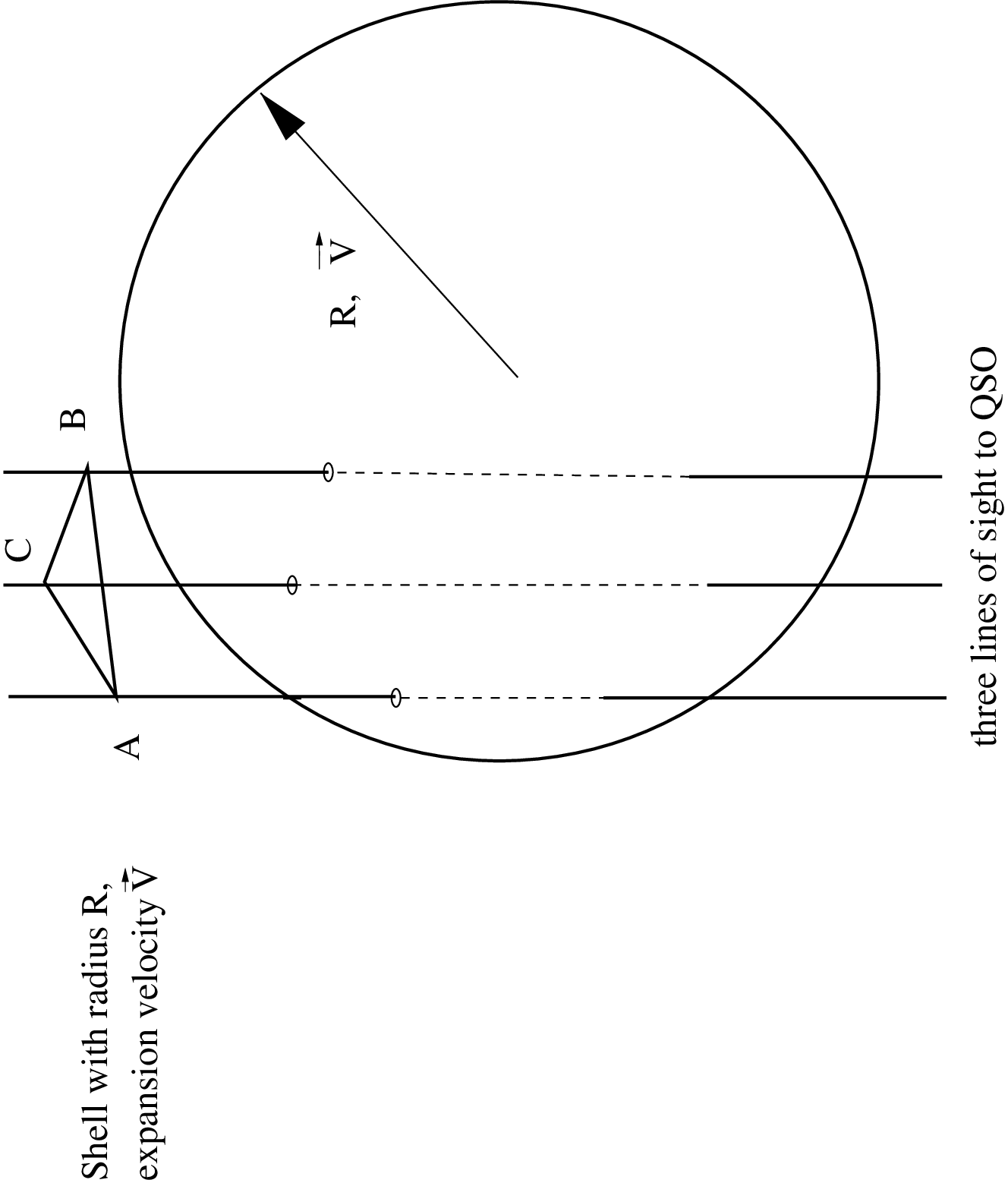}{4cm}{-90.}{28}{28}{-10}{240}
\vskip-1.5cm
\caption{ 
\small left: MgII structure in the three lines of sight ($A, B, C$) to Q2237+0305,
Right: bubble model, showing the possible geometry.}
\end{figure} 

\section{Conclusions}

We have argued that evidence for gasdynamical processes at high
redshift, including the effects of stellar feedback on the
intergalactic medium can be found (or at least sought) by probing space
with multiple closely spaced lines of sight to gravitationally lensed
QSOs. There is little evidence for disturbances in the low density
Lyman $\alpha$ forest, and simple models indicate that it may be hard
to fill a dominant fraction of the volume of the universe with winds
and not notice them as small scale disturbances in the Ly$\alpha$
forest.

However, the saturated Ly$\alpha$ systems and high ionization metal
absorption lines in general do show signs of energy input on scales
down to a few hundred parsecs. The degree of difference between the
lines of sight allows us to estimate how long ago the energy input must
have occured. The CIV gas in particular appears to be loosely
associated with galaxies, and seems to repeatedly have been influenced
by galactic feedback.

Relatively little is known about the small scale properties of low
ionization gas, but the few systems studied are consistent with having
structure at least down to a scale of 10 pc, where CIV gas clouds are
almost always homogeneous. It appears that we are seeing gasdynamical
processes in the interstellar medium or gas bubbles that have been
blown off their parent galaxies and may be sweeping up matter from the ambient
intergalactic medium.

\medskip

I thank Wal Sargent, Tom Barlow,
and Bob Carswell for allowing me to present our results at this conference, and Bob
Carswell again for useful comments on the manuscript.

\end{document}